%% file: main.tex
\documentclass[10pt,conference]{IEEEtran} 
\IEEEoverridecommandlockouts
\usepackage{cite}
\usepackage{amsmath,amssymb,amsfonts}
\usepackage{algorithmic}
\usepackage{graphicx}
\usepackage{subcaption}
\usepackage{textcomp}
\usepackage[x11names]{xcolor}
\usepackage{xspace}
\usepackage{url}
\usepackage[most]{tcolorbox}
\usepackage{hyperref} %
\usepackage[T1]{fontenc} %
\usepackage{flushend}

\def\BibTeX{{\rm B\kern-.05em{\sc i\kern-.025em b}\kern-.08em
    T\kern-.1667em\lower.7ex\hbox{E}\kern-.125emX}}
\begin{document}
\newcommand{\todo}[1]{\textbf{\textcolor{red}{ToDO: #1}}}
\newcommand{\wei}[1]{\textcolor{orange}{wei: #1}}
\newcommand{\dao}[1]{{\footnotesize{\textcolor{blue}{[Dao: {#1}]}}}\xspace}
\newcommand{\fixme}[1]{\textcolor{red}{#1}}
\newcommand{\cameraready}[1]{\textcolor{blue}{#1}}
\newcommand{\approach}{iAudit\xspace}
\newcommand{\tool}{iAudit\xspace}
\newcommand{\name}{iAudit\xspace}
\newcommand{\recheck}[1]{\textcolor{green}{ #1}}
\newcommand{\revision}[1]{\textcolor{blue}{ #1}}

\newcommand{\myfig}{Fig.\xspace}
\newcommand{\mysec}{Sec.\xspace}

\def\figref#1{Fig.~\ref{fig:#1}}
\def\figlabel#1{\label{fig:#1}\label{p:#1}}
\def\tabref#1{Table~\ref{tab:#1}}
\def\tabsref#1{Tables~\ref{tab:#1}}
\def\tablabel#1{\label{tab:#1}\label{p:#1}}
\def\Secref#1{Sec.~\ref{sec:#1}}
\def\secref#1{Sec.~\ref{sec:#1}}
\def\seclabel#1{\label{sec:#1}}
\def\qref#1{Eq.~\ref{eqn:#1}}
\def\eqrefn#1{\ref{eqn:#1}}
\def\eqsref#1#2{Eqs.~\ref{eqn:#1}/\ref{eqn:#2}}
\def\eqlabel#1{\label{eqn:#1}}
\def\subsp#1{P_{\mbox{{\scriptsize\rm #1}}}}
\def\appref#1{Appendix~\ref{sec:#1}}

\newcommand{\answer}[2]{
  \begin{tcolorbox}[enhanced, left=3mm,right=3mm,
    colback=gray!10, colframe=gray!80, boxrule=0pt,
    borderline west={4pt}{0pt}{gray!90},
    ]
    \textbf{Answer for RQ#1:}
    #2
    \end{tcolorbox}
}

\title{Combining Fine-tuning and LLM-based Agents for Intuitive Smart Contract Auditing with Justifications}

\author{
\IEEEauthorblockN{Wei Ma$^{1}$,
	 Daoyuan Wu$^{2}$\textsuperscript{\IEEEauthorrefmark{1}}, 
	 Yuqiang Sun$^{1}$, 
	 Tianwen Wang$^{3}$, 
	 Shangqing Liu$^{1}$,
    Jian Zhang$^{1}$,
	 Yue Xue$^{4}$, 
	 Yang Liu$^{1}$}
\IEEEauthorblockA{
\textit{$^1$ Nanyang Technological University, Singapore, Singapore} \\
\textit{$^2$ The Hong Kong University of Science and Technology, Hong Kong SAR, China} \\
\textit{$^3$ National University of Singapore, Singapore, Singapore} \\
\textit{$^4$ MetaTrust Labs, Singapore, Singapore} \\
ma\_wei@ntu.edu.sg, daoyuan@cse.ust.hk, suny0056@e.ntu.edu.sg, tianwenw.vk@gmail.com, \\liu.shangqing@ntu.edu.sg, jian\_zhang@ntu.edu.sg, xueyue@metatrust.io, yangliu@ntu.edu.sg
}
}

\maketitle
\begingroup\renewcommand\thefootnote{\IEEEauthorrefmark{1}}
\footnotetext{Corresponding author: Daoyuan Wu. Work conducted while at NTU.}
\endgroup

\input{sections/abstract}

\input{sections/introduction}

\input{sections/background}
\input{sections/approach}
\input{sections/evaluate}

\input{sections/related}
\input{sections/threats}
\input{sections/limitation}

\input{sections/conclusion}

\section*{Acknowledgment}
We thank all the reviewers for their detailed and constructive comments.
We also express our gratitude to all colleagues at MetaTrust Labs for their assistance in deploying the TrustLLM's iAudit model.
This research/project is supported by the National Research Foundation, Singapore, and the Cyber Security Agency under its National Cybersecurity R\&D Programme (NCRP25-P04-TAICeN), the National Research Foundation, Singapore, and DSO National Laboratories under the AI Singapore Programme (AISG Award No: AISG2-GC-2023-008), and NRF Investigatorship NRF-NRFI06-2020-0001. Any opinions, findings and conclusions or recommendations expressed in this material are those of the author(s) and do not reflect the views of National Research Foundation, Singapore and Cyber Security Agency of Singapore.
Daoyuan Wu was also partially supported by an HKUST grant.

\bibliographystyle{IEEEtran}
\bibliography{reference}

\end{document}

%% file: sections/abstract.tex
\begin{abstract}
Smart contracts are decentralized applications built atop blockchains like Ethereum.
Recent research has shown that large language models (LLMs) have potential in auditing smart contracts, but the state-of-the-art indicates that even GPT-4 can achieve only 30\% precision (when both decision and justification are correct).
This is likely because off-the-shelf LLMs were primarily pre-trained on a general text/code corpus and not fine-tuned on the specific domain of Solidity smart contract auditing.

In this paper, we propose \tool, a general framework that combines fine-tuning and LLM-based agents for intuitive smart contract auditing with justifications.
Specifically, \tool is inspired by the observation that expert human auditors first perceive what could be wrong and then perform a detailed analysis of the code to identify the cause.
As such, \tool employs a two-stage fine-tuning approach: it first tunes a Detector model to make decisions and then tunes a Reasoner model to generate causes of vulnerabilities. %
However, fine-tuning alone faces challenges in accurately identifying the optimal cause of a vulnerability. %
Therefore, we introduce two LLM-based agents, the Ranker and Critic, to iteratively select and debate the most suitable cause of vulnerability based on the output of the fine-tuned Reasoner model.
To evaluate \tool, we collected a balanced dataset with 1,734 positive and 1,810 negative samples to fine-tune \tool.
We then compared it with traditional fine-tuned models (CodeBERT, GraphCodeBERT, CodeT5, and UnixCoder) as well as prompt learning-based LLMs (GPT4, GPT-3.5, and CodeLlama-13b/34b).
On a dataset of 263 real smart contract vulnerabilities, \tool achieves an F1 score of 91.21\% and an accuracy of 91.11\%.
The causes generated by \tool achieved a consistency of about 38\% compared to the ground truth causes.
\end{abstract}

%% file: sections/introduction.tex
\definecolor{introboxcolor}{HTML}{303285}

\newcommand{\introbox}[1]{
  \begin{tcolorbox}[enhanced, left=3mm,right=3mm,
    colback=introboxcolor!10, colframe=introboxcolor!80, boxrule=0pt,
    borderline west={4pt}{0pt}{introboxcolor!90},
    ]
    #1
    \end{tcolorbox}
}

\introbox{
    ``One of the big skills in bug bounties that's really difficult to teach is intuition. Everything I do I am following my intuition. It's what looks interesting and what doesn't look right.''
    
    \hfill --- Katie Paxton-Fear
    
    \hfill One of the million-dollar-earning hackers~\cite{whitneyGooglePaidOut2024}.
}

\section{Introduction}
\label{sec:introduction}

Smart contracts have emerged as a key application based on blockchain technology since the advent of Ethereum.
Due to their openness, transparency, and irreversibility, smart contracts have become the foundation of decentralized financial applications (DeFi).
However, since DeFi manages a significant amount of digital assets, identifying and fixing vulnerabilities in smart contracts is crucial.
Currently, the real vulnerabilities exploited by hackers in smart contracts are mainly due to logical flaws~\cite{zhangDemystifyingExploitableBugs2023b}, which render traditional pattern-based program analysis~\cite{feistSlitherStaticAnalysis2019,fangSomo2023,Brent_Grech_Lagouvardos_Scholz_Smaragdakis_2020,Chen_Xia_Lo_Grundy_Luo_Chen_2022,Brent_Jurisevic_Kong_Liu_Gauthier_Gramoli_Holz_Scholz_2018,ZEUS2018,Securify2018,Mossberg_Manzano_Hennenfent_Groce_Grieco_Feist_Brunson_Dinaburg_2019} less effective.
According to Defillama Hacks~\cite{defillama}, vulnerability attacks have caused losses of around \$7.69 billion as of March 2024.
Hence, there is an urgent need for innovative methods to combat these emerging threats.

Recent research~\cite{davidYouStillNeed2023c,sunGPTScanDetectingLogic2023c,huLargeLanguageModelPowered2023b,sunLLM4VulnUnifiedEvaluation2024a} has shown that large language models (LLMs) have potential in auditing smart contracts, especially in demonstrating superior performance in detecting logic vulnerabilities~\cite{zhangDemystifyingExploitableBugs2023b,sunGPTScanDetectingLogic2023c}.
However, a recent systematic evaluation study~\cite{sunLLM4VulnUnifiedEvaluation2024a} shows that even when equipping the LLM-based vulnerability detection paradigm with a state-of-the-art approach, namely enhancing GPT-4 with summarized vulnerability knowledge in a Retrieval Augmented Generation (RAG)~\cite{lewisRetrievalaugmentedGenerationKnowledgeintensive2020} fashion, it still achieves only $\sim$30\% precision when both the decision (i.e., whether the subject code is vulnerable) and justification (i.e., pinpointing the correct vulnerability type) are correct.
This can be attributed to the fact that off-the-shelf LLMs (e.g., GPT-4), which were primarily pre-trained on a general text/code corpus, were not fine-tuned for the specific domain of Solidity\footnote{Solidity is a mainstream language for smart contract development.} smart contract auditing.

Fine-tuning~\cite{tianFinetuningLanguageModels2023,lvFullParameterFinetuning2023} could be a promising approach to embed Solidity-specific vulnerability data into the model itself, compared to RAG~\cite{balaguerRAGVsFinetuning2024}, and thus address the problem mentioned above.
In particular, by fine-tuning an LLM with vulnerable and non-vulnerable code, it could effectively \textit{perceive} whether a new piece of code is vulnerable or not.
According to insights from a million-dollar-earning hacker mentioned in the prologue, such intuition is quite important for vulnerability auditing.
As such, instead of fine-tuning a single model to generate both vulnerability decisions (i.e., Yes or No) and the causes of vulnerabilities (i.e., the type or reason) simultaneously, we propose a novel two-stage fine-tuning approach.
This approach first tunes a Detector model to make decisions only, and then tunes a Reasoner model to generate the causes of vulnerabilities.
In this way, the fine-tuned LLMs could mimic human hackers by first making intuitive judgments and then performing follow-up analysis of the code to identify the reasons for vulnerabilities.

We implement this ``perception-then-analysis'' fune-tuning into a general framework called \name\footnote{\name is deployed as an auditing module of MetaTrust Labs' TrustLLM; see \url{https://huggingface.co/MetaTrustSig}.} for intuitive smart contract auditing.
In this implementation, \name allows Detector to make multiple intuitive judgments, each representing one perception.
To achieve this, \name generates multiple variant prompts for the same vulnerability label to tune Detector and similarly employs multiple variant prompts for the same vulnerability reason to tune Reasoner.
While it is possible to determine the optimal decision based on majority voting, fine-tuning alone cannot \textit{identify the optimal cause for a vulnerability during the inference phase}.
To address this new problem, we introduce the concept of LLM-based agents to the paradigm of fine-tuning in \name.
Specifically, we introduce two dedicated LLM-based agents, the Ranker and Critic agents, to iteratively select and debate the most appropriate cause of vulnerability based on the output of the fine-tuned Reasoner model.

To obtain high-quality data for training and testing \name, we propose leveraging reputable auditing reports to collect positive samples and employing our own data enhancement method to derive negative samples.
Eventually, we collected a balanced dataset consisting of 1,734 positive samples, i.e., vulnerable functions with reasons from 263 smart contract auditing reports, and 1,810 negative samples, i.e., non-vulnerable benign code.
We then compared \name with traditional full-model fine-tuning methods, including CodeBERT, GraphCodeBERT, CodeT5, and UnixCoder, as well as with prompt learning-based LLMs, such as GPT-4/GPT-3.5 and CodeLlama-13b/34b.
Our experimental results show that \name achieved an F1 score of 91.21\%, significantly outperforming prompt learning-based LLMs (which are in the range of 60\%+) and also notably beating other fine-tuned models (which are in the range of 80\%+) that used the same training data as ours.
Furthermore, in terms of alignment with ground-truth explanations, \name's output is clearly superior to that of other models, reaching a consistency rate of 37.99\%.
In contrast, the second-ranked GPT-4 achieves only 24\%.

Besides the evaluation results, we also conducted three ablation studies to further justify \name's two-stage fine-tuning and majority voting strategies, as well as to measure the impact of additional call graph information on the model's performance.
We summarize the key findings as follows:
\begin{itemize}
    \item \name's two-stage approach achieved better detection performance than the integration model, which outputs labels and reasons simultaneously. We also experimentally confirmed that the model struggles to focus on the labels when required to output both types of information.

    \item Majority voting enhances the detection performance and stability. Using multiple prompts also allows the model to perform better than when using a single prompt.

    \item Call graph information may enable the model to make better judgments in some cases, but we also observed situations where this additional information could potentially confuse the model, thereby reducing its performance.
\end{itemize}

\textbf{Roadmap.}
The rest of this paper is organized as follows.
We first introduce the relevant background in \secref{background}, followed by the design of \name in \secref{approach}.
We then present our experimental setup and the results in \secref{evaluation}.
After that, we discuss related work and the limitations in \secref{related} and \secref{threats}, respectively.
Finally, \secref{conclusion} concludes this paper.

\textbf{Availability.}
To facilitate future research and comparison, we have made the inference code and dataset available at~\cite{iAudit}.

%% file: sections/background.tex
\section{Background}
\label{sec:background}

\subsection{Pre-trained Models and Large Language Models}

Pre-trained models are models that have been initially trained on large datasets.
These models can be quickly adapted to various specific tasks with minimal adjustments, avoiding the complex training process from scratch.
Currently, most pre-trained models adopt an architecture based on transformers~\cite{vaswani2017attention}.
The innovation of this approach is that pre-trained models leverage large data and well-designed tasks for effective feature learning, which has been proven effective in multiple fields, such as text processing, image recognition, and software engineering.
The standard transformer structure consists of one encoder and one decoder, which are structurally similar but function differently.
Pre-trained models can be classified into encoder-based, decoder-based, or encoder-decoder combined types depending on the transformer structure used.
For example, encoder-based models are represented by BERT~\cite{devlin2018bert} and CodeBERT~\cite{feng-etal-2020-codebert}, decoder models by the GPT series~\cite{radford2019language,brown2020language}, and encoder-decoder models by BART~\cite{lewis2019bart}, T5~\cite{10.5555/3455716.3455856}, and CodeT5~\cite{wang2021codet5}.
Compared with general pre-trained models, Large Language Models (LLMs)~\cite{zhao2023survey,10.1145/3641289} differ significantly in their used larger data and model scales.
These models are trained by learning world-wide knowledge bases, typically reaching billions in scale.
As the model size, data volume, and computational capacity increase, performance also improves, as revealed by the Scaling Laws~\cite{kaplan2020scaling}.
Closed-source LLMs like GPT-3.5, GPT-4, and Gemini~\cite{google} offer their services externally through APIs, while open-source models like Llama2~\cite{touvron2023llama} can achieve performance comparable or better to closed-source models after fine-tuning.

\subsection{Parameter-Efficient Fine-Tuning}

LLMs have extremely large parameters.
Fully fine-tuning a large language model requires significant hardware resources and is very costly.
Therefore, lightweight parameter fine-tuning~\cite{xu2023parameter,wan2023efficient} is currently the main method of using LLMs compared to fully fine-tuning them.
Although LLMs can be used without task-specific fine-tuning through in-context learning~\cite{dong2022survey}, this usually requires carefully prepared prompts.
Furthermore, research has found that partial fine-tuning of LLMs with smaller parameters can achieve or even surpass the effects of huge models~\cite{hu-etal-2023-llm,song2023sparse}.
These fine-tuning methods differ from full-model fine-tuning by focusing only on fine-tuning additional parameters while keeping the large model weights fixed, known collectively as parameter-efficient fine-tuning~\cite{xu2023parameter,wan2023efficient}.
They can be generally categorized into four types: Adapter~\cite{hu-etal-2023-llm}, Low-Rank Adaptation~(LoRA)~\cite{hu2021lora}, prefix tuning~\cite{li2021prefix}, and prompt tuning~\cite{lester-etal-2021-power}.

Adapter~\cite{hu-etal-2023-llm} adds a lightweight additional module to each layer of the model to capture information specific to downstream tasks.
During optimization, only the parameters of the additional module are optimized.
Since the number of parameters in the Adapter is much smaller than that of the model itself, it significantly reduces the overall parameter count and computational complexity of the model.

Low-Rank Adaptation (LoRA)~\cite{hu2021lora} is a parameter-efficient adaptation method for LLMs, which adjusts LLMs for downstream tasks at a lower parameter cost.
The core idea of LoRA is to introduce additional, low-rank adaptation parameters into the self-attention mechanism, effectively adjusting the model to suit new tasks with minimal addition of extra parameters.

Prefix tuning~\cite{li2021prefix} adds a ``prefix'' sequence to each layer of the model, serving as additional context input.
This method allows the model to adapt to specific tasks while retaining most of the knowledge acquired during pre-training.
Unlike prefix tuning, prompt tuning~\cite{lester-etal-2021-power} adds prompt tokens to the input, which can be placed at any position. 

To sum up, using adapters can increase inference latency~\cite{hu2021lora,mundra2024comprehensive}.
Prefix or prompt tuning is subject to structural constraints that inhibit the learning of new attention patterns~\cite{petrov2024when}.
LoRA is an efficient method with low cost and can have a performance close to the full fine-tuning approach~\cite{hu2021lora}. %

\subsection{Smart Contracts and Their Vulnerabilities}

Smart contracts are essential for realizing decentralized finance~\cite{zetzsche2020decentralized} as an application layer of blockchain technology. 
According to data from DeFiLlama~\cite{defillama_chain}, as of March 2024, the total value locked in the top three blockchain platforms (Ethereum, Tron, and BSC) has reached \$73 billions.
Given the close relationship between smart contracts and economic interests, their security has attracted widespread attention.
Vulnerabilities in smart contracts can lead to significant losses, such reentrancy attacks and access-control attacks~\cite{praitheeshan2019security}. 

In the real world, hackers employ even more complex tactics.
Currently, to address vulnerabilities in smart contracts, various static and dynamic detection tools~\cite{feistSlitherStaticAnalysis2019,fangSomo2023,Brent_Grech_Lagouvardos_Scholz_Smaragdakis_2020,Chen_Xia_Lo_Grundy_Luo_Chen_2022,Brent_Jurisevic_Kong_Liu_Gauthier_Gramoli_Holz_Scholz_2018,ZEUS2018,Securify2018,Mossberg_Manzano_Hennenfent_Groce_Grieco_Feist_Brunson_Dinaburg_2019} are used to test contract security.
Unfortunately, some complex vulnerabilities are hard to be found by these detection tools.
For example, in a sandwich attack~\cite{zust2021analyzing}, attackers monitor other pending transactions and execute their transactions first upon spotting a high-value yet uncompleted transaction.
Due to this preemptive action, the attack transaction will be executed at a higher price, allowing the attacker to immediately sell the acquired excess profit for profit.
Many well-funded project teams also invite third parties to audit their smart contracts before public release to ensure their safety.

%% file: sections/approach.tex
\begin{figure*}[t!]
    \centering
    \includegraphics[width=1\textwidth]{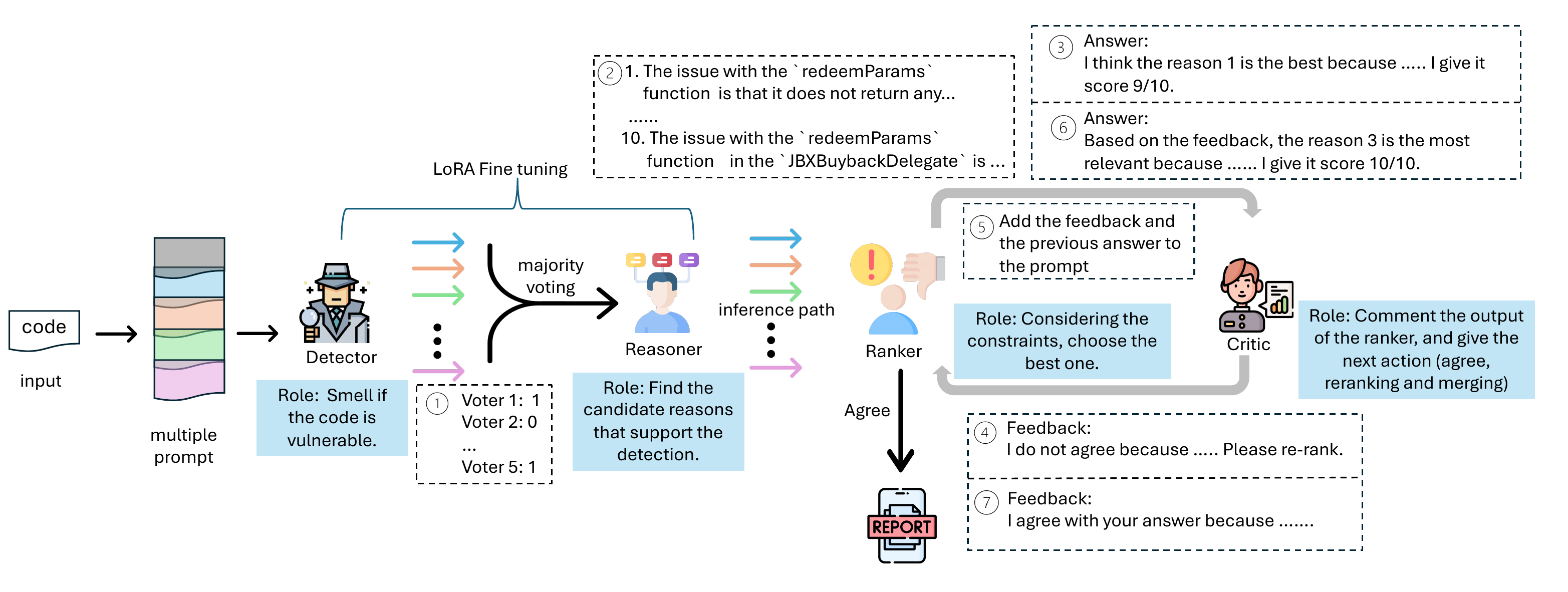}
    \caption{An overview of \approach, featuring its four roles: Detector, Reasoner, Ranker, and Critic.}%
    \label{fig:overview}
\end{figure*}

\section{Design of \name}
\label{sec:approach}
As motivated in \mysec\ref{sec:introduction}, \name employs a novel two-stage fine-tuning approach and combines it with LLM-based agents for intuitive smart contract auditing with justifications.
As shown in \myfig~\ref{fig:overview}, \name has the following four roles:
\begin{itemize}
    \item \textbf{Detector} is the key component for achieving intuitive smart contract auditing.
    By fine-tuning an LLM with vulnerable and non-vulnerable code, Detector can discern whether a piece of code is vulnerable, much like how a human hacker perceives a potential vulnerability. %
    
    \item \textbf{Reasoner} takes the initial vulnerability perception from Detector to further analyze the potential causes of the vulnerability based on Detector's decision.
    By connecting Detector's output with Reasoner's reasoning during both training and inference, \name achieves two-stage fine-tuning. 
    
    \item To identify the optimal cause of a vulnerability during the inference phase, we further introduce the concept of LLM-based agents into the fine-tuning paradigm in \name.
    Specifically, \textbf{Ranker} evaluates the reasons for each potential vulnerability, selecting a top explanation, while \textbf{Critic} further assesses Ranker's output to debate and determine the most appropriate cause of the vulnerability.
\end{itemize}

\textbf{Challenges.}
While \name's four roles in \myfig~\ref{fig:overview} are intuitive, training and coordinating them well for effective smart contract auditing with reasonable justifications is difficult.
More specifically, we encountered the following four challenges during the design and implementation of \name:
\begin{description}
    \item [C1:] \textit{How to collect and derive high-quality training data?}
    For a fine-tuned model like \name, obtaining high-quality training data is always crucial.
    We propose leveraging reputable auditing reports to collect positive samples and employing our own data enhancement method to derive negative samples.
    Since this part is independent of \name's design, we defer its presentation to the end of this section in \mysec\ref{sec:data}.

    \item [C2:] \textit{How to make effective vulnerability judgements?}
    While fine-tuning a model with vulnerable and non-vulnerable code is straightforward, tuning it to be effective with limited data presents a challenge.
    We make an effort towards addressing this problem in \mysec\ref{sec:detector} by opting to use multiple prompts for fine-tuning rather than a single prompt.
    The advantages of this approach are twofold: (i) it enriches the training dataset by increasing the volume of data, and (ii) it diminishes the bias associated with a single prompt, thereby enhancing the reliability of the results~\cite{zhou-etal-2022-prompt}.
    Optimal vulnerability perception could thus be achieved through majority voting.
    \item [C3:] \textit{How to effectively connect Detector's vulnerability sensing with Reasoner's vulnerability reasoning?}
    The fine-tuning of \name is unique because it employs a two-stage fine-tuning approach with the Detector and Reasoner models.
    Therefore, how to effectively connect these two models becomes a new issue not encountered in traditional fine-tuning.
    We present this aspect of \name's design in \mysec\ref{sec:reasoner}.

    \item [C4:] \textit{How to obtain the optimal vulnerability cause from Reasoner's output?}
    Since Reasoner also employs multi-prompt fine-tuning, it is necessary to identify the optimal cause of vulnerability among the multiple causes output by Reasoner.
    We introduce two LLM-based agents, namely the Ranker and Critic components, in \mysec\ref{sec:rankerANDcritic}, to iteratively select and debate the most appropriate cause of vulnerability.
\end{description}

\textbf{An Example of Workflow.}
To wrap up, \figref{overview} also illustrates an example of \tool's workflow.
Initially, Detector perceives code vulnerabilities using five different inference paths (prompts).
The perceived results are then subjected to majority voting to determine a consensus label.
Based on the voting result, Reasoner interprets this outcome according to different inference paths, resulting in ten answers (each considering the context of the code location or not).
Next, Ranker selects Reason 1 with a confidence score 9/10 and explains this choice.
Critic challenges this choice and advises Ranker to re-evaluate.
Taking Critic's feedback into account, Ranker re-ranks the ten reasons and selects Reason 3 with a confidence score of 10/10.
Critic reviews Ranker's choice again and agrees with this decision.
The loop is completed, and the final reason is returned to the user.

\subsection{Using Multi-prompt Tuning and Majority Voting for Effective Vulnerability Judgements in the Detector}
\label{sec:detector}

Detector is a fine-tuned expert model responsible for assessing whether code poses any risk.
It mimics human intuitive judgment upon seeing a piece of code, assessing whether there are any issues.
We employed LoRA~\cite{hu2021lora} to fine-tune CodeLlama-13b~\cite{roziere2023code} in the instruction manner~\cite{alpaca} based on a high quality of dataset.
During training, for the same input code, we wrap it with multiple prompts.
These prompts, with different instruction formats, represent the different inference paths, as illustrated in \myfig\ref{fig:overview}.
In the inference phase of Detector, based on the output results of each prompt, we adopt a majority voting approach to determine the input label and use the voting ratio as the confidence score.
Based on Detector's majority voting result, Reasoner in \mysec\ref{sec:reasoner} then generates different reasons according to different inference paths.

It is worth noting that for the choice of the base model, we randomly selected 16 real logical vulnerabilities to evaluate three popular open-source models: StarCoder, Llama2, and CodeLlama. Upon manual review, we found that StarCoder sometimes refused to respond, and Llama2 provided one contradictory response with two labels. In contrast, CodeLlama offered a more stable response, which led us to choose CodeLlama as the foundational model for further fine-tuning.

\begin{figure}[t!]
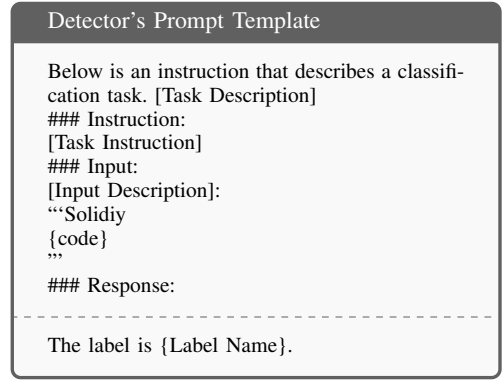

\begin{center}
\scalebox{0.9}{
\begin{tcolorbox}[width=0.4\textwidth, colback=gray!5!white,colframe=gray!75!black,title=Detector's Prompt Template]
\small
  Below is an instruction that describes a classification task.
\small
[Task Description] 

\#\#\# Instruction:

[Task Instruction] 

\#\#\# Input: 

[Input Description]: \\
```Solidiy \\
\{code\} \\
''' \\
\#\#\# \small Response:
\tcblower
\small
The label is \{Label Name\}.
 
\end{tcolorbox}}
\end{center}
\caption{The Prompt Template Used by Detector.}
\label{fig:detector_prompt_template}
\end{figure}

The prompt template used by Detector is demonstrated in \figref{detector_prompt_template}.
Above the dashed line is the input $x$ for our model. ``\{code\}'' is the placeholder for the input code.
Below the dashed line, ``The label is  \{Label Name\}'' is our target training output $y$, with ``\{Label Name\}'' being the label placeholder, which can be either ``safe'' and ``vulnerable.''
The left table~(Detector's Multiple Prompts) in \figref{promtpts_reasoner_detector} details the [Task Description], [Task Instruction], and [Input Description], listed as notations \textit{a}, \textit{b}, and \textit{c}, respectively.
We fine-tune CodeLlama-13b using LoRA in a \textit{generative} manner, as shown in Eq.~\ref{eq:clm_loss},
\begin{equation}\label{eq:clm_loss}
L(\theta) = -\sum_{t=1}^{T} \log P_{\theta}(y_t | x, y_{<t})  
\end{equation}
where $\theta$ represents the LoRA trainable parameters, $T$ is the output sequence length, and $P_{\theta}(y_t | x, y_{<t})$ is the probability of the model with parameters \( \theta \) generating a token \( y_t \) given the context \( x \) and all previous tokens \( y_{<t} \).
In this generative approach, the output at the time step $t$ is conditioned only on the previous time steps ($<t$).

After fine-tuning, during inference, the proposed input follows the training format, and we need to extract labels from the output via keyword matching, using ``safe'' and ``vulnerable.''
Since we employ multiple prompts, we obtain multiple label predictions and use majority voting to decide the final predicted label.
In majority voting, each inference path casts a vote for one of the available labels, ``safe'' or ``vulnerable,'' denoted as $l_0$ and $l_1$, respectively.
The label that receives more than half of the total votes is declared the winner.
Let $L = \{l_0, l_1\}$ represent the set of labels, and $V = \{v_1, v_2, \ldots, v_m\}$ represent the set of the prompt voter.
Each prompt voter $v_i$ casts a vote for one label.
The winning label $l_i$ is the one of $l_0$ and $l_1$ for which the following condition holds: $|\{v \in V : \text{vote}(v) = l_i\}| > \frac{m}{2}$.
This condition asserts that the winning label $o_w$ must receive more than half of the total votes $m$.
We also use the voting ratio of the winning label as the confidence score for the final decision.

\subsection{Connecting Detector for Reasoner's Tuning \& Inference}
\label{sec:reasoner}

Reasoner is an expert model responsible for reasoning about and explaining code vulnerabilities.
It interprets the majority voting result of the Detector, generating multiple alternative explanations.
During the Reasoner's LoRA fine-tuning process, our inputs include the code, its context, corresponding labels, and we construct zero-shot chain-of-thought (CoT)~\cite{kojima2022large} prompts with different command formats for training.
In the inference phase, Reasoner outputs multiple explanations based on the majority-voted label from Detector.
We constructed two types of prompts: the first type includes the label, code information, and its function call relationships; the second type includes only the label and code information.
In \mysec\ref{sec:evaluation}, we will investigate the impact of including function call relationships or not.
For each type, we designed five different instruction formats for the prompts, totaling ten inference paths, as illustrated in \myfig\ref{fig:overview}.

\begin{figure}[t!]
\begin{center}
\scalebox{0.9}{
\begin{tcolorbox}
[width=0.4\textwidth,colback=gray!5!white,colframe=gray!75!black,title=Reasoner's Prompt Template]
\small
Below is an instruction that describes a reasoning task.

[Task Description] 

\#\#\# Instruction:
[Task Instruction] 

\#\#\# Input: 

[Input Description]: \\
```Solidiy \\
\{code\} \\
``` \\
\#\#\# As a Caller: \textcolor{gray}{(Optional)}

[Caller Description]
```\\
\{caller info\} \\
```\\
\#\#\# As a Callee: \textcolor{gray}{(Optional)}

[Callee Description]

```\\
\{callee info\} \\
```\\
\#\#\# Response:

[Label Information + Zero-shot-CoT Tip]
\tcblower

\{Target Reason\}
 
\end{tcolorbox}}
\end{center}
\caption{The Prompt Template Used by Reasoner.}
\label{fig:reasoner_tempate_prompt}
\end{figure}

\begin{figure*}
    \centering
\includegraphics[width=1\textwidth]{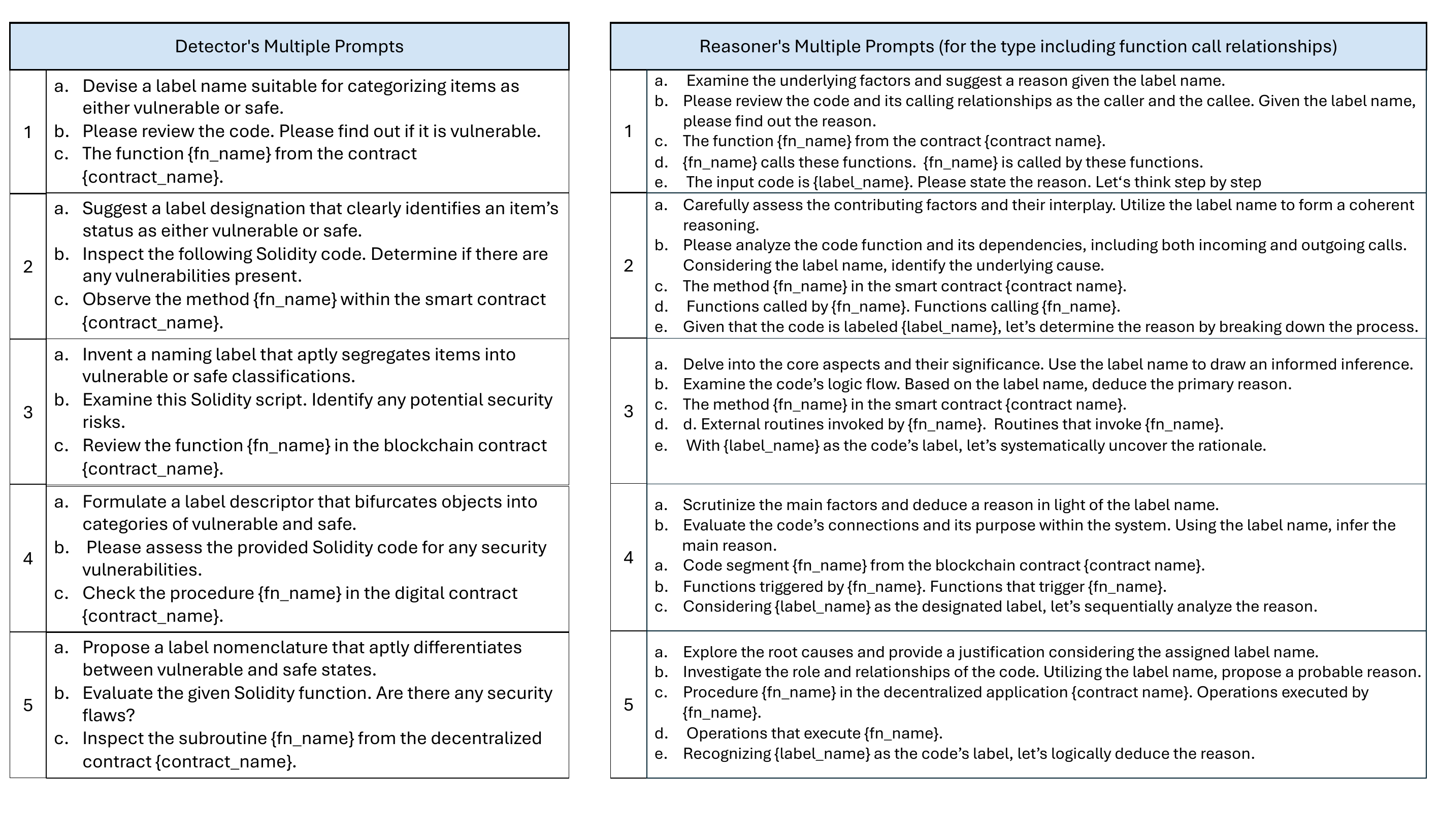}
    \caption{Detailed Multiple Prompts for Detector and Reasoner.}
    \label{fig:promtpts_reasoner_detector}
\end{figure*}

The prompt template used by the Reasoner is shown in \figref{reasoner_tempate_prompt}.
``{code}'', ``{caller info}'', ``{callee info}'', and ``{Target Reason}'' are placeholders for the input code, caller context, callee context, and the target output, respectively.
The right table~(Reasoner's Multiple Prompts) in \figref{promtpts_reasoner_detector} details the first prompt type with calling context, including [Task Description] denoted as \textit{a}, [Task Instruction] denoted as \textit{b}, [Input Description] denoted as \textit{c}, [Caller Description] and [Callee Description] denoted as \textit{d}, and [Label Information + Zero-shot-CoT Tip] denoted as \textit{e}.
For the second prompt type, the calling context is omitted.
Reasoner employed the same fine-tuning method as Detector, as shown by Eq.~\ref{eq:clm_loss}.
During inference, the proposed input prompt follows the training format, and Reasoner generates ten answers to interpret Detector's assessment.

\subsection{Ranking and Debating the Optimal Vulnerability Cause}
\label{sec:rankerANDcritic}

Ranker and Critic are two LLM-based agents collaborating to select the most appropriate cause of vulnerability from multiple explanations returned by Reasoner for a given code function.
Ranker performs two actions: ``\textit{rank}'' and ``\textit{merge}''.
``\textit{Rank}'' involves selecting the best explanation from the ones provided, while ``\textit{merge}'' involves integrating multiple selected explanations.
We define 10 constraints for Ranker to select the top explanation.
Critic evaluates Ranker's answer in conjunction with the code function, providing three next-step action instructions: ``\textit{agree}'', ``\textit{rerank}'', and ``\textit{merge}''.
``\textit{Agree}'' means the current answer is reasonable and can be returned to the user.
``\textit{Rerank}'' indicates that Ranker needs to re-select, considering Critic's feedback and previous answers.
``\textit{Merge}'' suggests that the top reasons provided must be integrated.

More specifically, Ranker employs the following 10 constraints in the prompt as its selection criteria:
\begin{enumerate}
    \item If one reason describes code that does not exist in the provided input, it is not valid.
    \item If one reason is not related to the code, the reason is not valid.
    \item If this reason violates the facts, the reason is unreasonable.
    \item If one reason is not related to the decision, the reason is not valid.
    \item If one reason assume any information that is not provided, the reason is not valid.
    \item If the code is safe and one reason supports the decision, please check if the code has other potential vulnerabilities. If the code has other potential vulnerabilities, the reason is not valid.
    \item The selected reason should be the most relevant to the decision.
    \item The selected reason must be the most reasonable and accurate one.
    \item The selected reason must be factual, logical and convincing.
    \item Do not make any assumption out of the given code.
\end{enumerate}

Both Ranker and Critic are LLMs agents implemented based on the Mixtral 8x7B-Instruct~\cite{jiang2024mixtral} model, the capability of which is close to that of larger LLMs~\cite{jiang2024mixtral,jiang2023mistral,xue2024openmoe}. Moreover, we have observed that the Mixture of Experts~(MoE)~\cite{jiang2024mixtral} model can more effectively output data in the predetermined format than other models, making it easier for us to handle the output.

\subsection{High-quality Training Data Collection and Enhancement}
\label{sec:data}

The quality of training data is crucial for fine-tuning LLMs.
To collect positive samples, namely risky vulnerability code, we can employ auditing reports from reputable industry companies, such as Trail of Bits, Code4rena, and Immunefi.
Specifically, we crawled and parsed 1,734 vulnerable functions with reasons from 263 smart contract auditing reports, which were assembled by a popular auditing website called Solodit~\cite{Solodit}.

However, to train our model, we also need non-vulnerable benign code (i.e., negative samples), but this type of data is missing in the audit reports.
Therefore, we propose our own data enhancement method to derive high-quality negative samples.
Specifically, we adopt the GPT-4-based approach described in LLM4Vuln~\cite{sunLLM4VulnUnifiedEvaluation2024a} to extract vulnerability knowledge from vulnerability reports on Code4rena.
This includes the functionality descriptions of vulnerable functions and the code-level reasons why the vulnerabilities occur.
We then cluster this raw vulnerability knowledge based on the functionality descriptions into groups using Affinity Propagation~\cite{frey2007clustering} as described in~\cite{yi2022empirical} and use GPT-4 to summarize a functionality description for each group.
With the hierarchical information of group functionality, individual functionality, and vulnerability negligence, we employ the hierarchical GPT-based matching (i.e., matching the group first, then matching functionality and negligence) in GPTScan~\cite{sunGPTScanDetectingLogic2023c} to obtain the label information for tested code.
A function is labeled as a negative sample if no vulnerability information matches.
All prompts used are from LLM4Vuln and GPTScan.

Eventually, we collected a balanced dataset with 1,734 positive samples and 1,810 negative samples.
In this dataset, vulnerable functions have a median of 49.5 lines of code and a complexity of 18.5, while safe functions have a median of 35.5 lines of code and a complexity of 13.5. The complexity distributions between the two types are somewhat similar but with a slight difference, indicated by a Kullback-Leibler (KL) divergence value of 0.1.
This dataset was divided into \textbf{training}, \textbf{validation}, and \textbf{test} subsets, containing \textbf{2,268}, \textbf{567}, and \textbf{709} entries, respectively.
During training, we use the training and validation sets. During testing, we use the test set.

\begin{figure}[t!]
    \centering
    \includegraphics[width=0.49\textwidth]{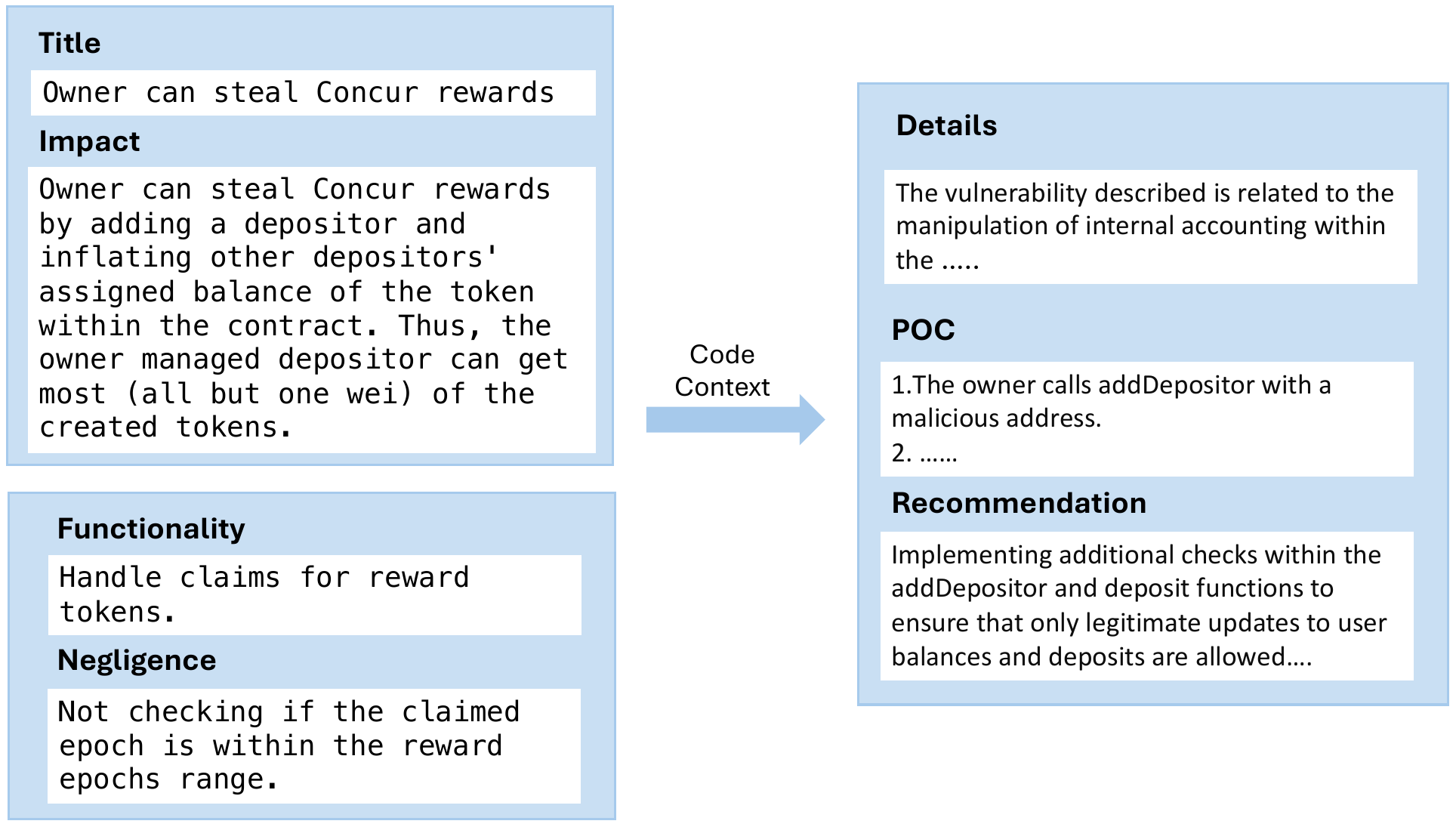}
    \caption{Data Enhancement for Expanding Vulnerability Explanations based on GPT-3.5.}
    \label{fig:example_data_enhanced}
\end{figure}

After collecting the labeled and unlabeled data, we also obtained corresponding explanations for the vulnerabilities.
However, the quality of these vulnerability justifications varies considerably.
Furthermore, some data contain external links, which may cause the model to hallucinate and output non-existent links.
To improve the interpretability of the reasons behind the vulnerabilities in the dataset, we used GPT-3.5 to enhance the existing explanations, expanding on the explanations of the vulnerabilities, proofs of concept (PoC), and recommended fixes.
\figref{example_data_enhanced} shows an example, where the left part presents the original reasons for the vulnerability, which are short and lack detail. We thus use them as prompts, along with the code context, to instruct GPT-3.5 to generate more detailed descriptions, including the PoC and the mitigation recommendation.

Note that we chose GPT-3.5 mainly because it achieves a good balance between cost and effectiveness—much cheaper than GPT-4, yet comparable in quality for this specific task during our manual comparison. We also added constraints to our prompts during this process to ensure that the enhanced explanations aligned with the actual vulnerabilities.

%% file: sections/evaluate.tex
\section{Evaluation}
\label{sec:evaluation}

\subsection{Experimental Setup}
\label{sec:setting}

In our study, we carefully selected a series of benchmark models, categorized into two groups: LLMs for zero-shot learning and pre-trained code models based on fine-tuning, to ensure a comprehensive and sound comparative analysis.
For zero-shot learning LLMs, we chose CodeLlama-13b-Instruct, CodeLlama-34b-Instruct~\cite{roziere2023code}, GPT-3.5, and GPT-4 as benchmarks, representing the current state-of-the-art.  %
Additionally, we selected CodeBERT~\cite{feng-etal-2020-codebert}, GraphCodeBERT~\cite{guo2020graphcodebert}, CodeT5~\cite{wang2021codet5}, UnixCoder~\cite{guo2022unixcoder}, and CodeLlama-13b~\cite{roziere2023code} to train classifiers.
Among these, CodeBERT, GraphCodeBERT, CodeT5, and UnixCoder underwent a complete model fine-tuning process to adapt to the specific code classification task.
In particular, CodeLlama-13b employs LoRA for lightweight tuning and uses the last token representation for classification.
Note that that our method is different; \tool's Detector achieves classification by generating label names as task outputs.

\subsection{Research Questions~(RQs)}
Since our proposed method comprises two core functions: vulnerability detection and reason explanation, we designed a series of experiments to evaluate and demonstrate the performance and effectiveness of both tasks. These experiments aim to answer the following research questions (RQs):

\paragraph{RQ1 - Performance Comparison} \textbf{How does the performance of \approach in detecting vulnerabilities compare to other models?}
This question aims to understand how the effectiveness of Detector in detecting vulnerabilities compares to that of other existing models. The focus is on comparative analysis, involving metrics accuracy, precision, recall, and F1 score, to evaluate and contrast the performance. 

\paragraph{RQ2 - Explanation Alignment} \textbf{To what extent do the explanations generated by \approach's Reasoner align with the real reasons?} RQ2 concerns the quality of the explanations the Reasoner provided for the decision of the Detector. It questions whether the reasons given by \approach correspond to the actual reasons behind the vulnerabilities, emphasizing the interpretability and trustworthiness of the model.

\paragraph{RQ3 - Two-stage Approach vs. An Integration Model} \textbf{How does \name compare with an integration model that performs detection and reasoning simultaneously?} Our method is based on a generative model, with two models trained on the generated labels and reasons, respectively. Another approach uses a single model to generate both reasons and labels. This question explores the effectiveness and impact of integrating the Detector and Reasoner components into one. %

\paragraph{RQ4 - Effectiveness of Majority Voting} \textbf{Can majority voting improve the effectiveness of the Detector?} RQ4 investigates whether the effectiveness of the Detector can be improved by adopting a majority voting mechanism. Majority voting, a technique that makes the final decision based on the majority output of multiple models, may improve the robustness and accuracy of the method.

\paragraph{RQ5 - Impact of Additional Information} 
The \textbf{call graph} illustrates the interaction of code with other components within the project, which is expected to be advantageous for our task. We address two specific research sub-questions: 
\begin{itemize}
    \item RQ5.1. Can the call graph enhance the Detector performance?
    \item RQ5.2. In what way does the call graph influence our explanation generation process, specifically within the Reasoner-Ranker-Critic pipeline?
\end{itemize}

Besides the RQs above, we also used our model to audit two bounty projects (currently anonymous) on Code4rena.
We invited audit experts to verify our findings.
In the end, we found 6 critical vulnerabilities, which were recognized by the project team or audit experts. 
In particular, one vulnerability was not discovered by any tools, marked as a great finding. This demonstrates the real-world value of \name.
Due to page limitations, we have included these case studies in the supplementary materials for interested readers.

\subsection{RQ1 - Performance Comparison}

Firstly, we compared \tool with LLMs based on zero-shot learning, as shown in \tabref{zero_llms}. Our method also uses a zero-shot approach during the inference phase. We considered two proprietary models (GPT-4 and GPT-3.5) and three open-source models (CodeLlama-13b and CodeLlama-34b). For the open-source models, we strictly adhered to their prompt formats. Huggingface Transformer~\cite{huggingface_transformer} has integrated these prompt formats into its framework. The format conversion is completed by calling \texttt{apply\_chat\_template}. CodeLlama requires adding [INST] and [/INST] as well as special tags <<SYS>> and <</SYS>>. 
As shown in \tabref{zero_llms}, after fine-tuning, our proposed strategy significantly outperforms the baseline models in the zero-shot scenario in terms of F1, accuracy, and precision, achieving high scores of $0.9121$, $0.8934$, and $0.9111$, respectively. However, when examining the recall, we notice that, the baseline models all performed excellently. Notably, GPT-4 and GPT-3.5 achieved a recall score of $1$. We checked the confusing matrix and found that all test data are labelled by the vulnerability. For GPT-4 and GPT-3.5, we adopted the prompts which are provided by our industrial partner, MetaTrust Labs, a Web3 security company. %

\begin{table}[t!]
\caption{Performance comparison between \approach's Detector and zero-shot LLMs.}
\label{tab:zero_llms}
\centering
\large
\scalebox{0.8}{
\begin{tabular}{l|c|c|c|c}
\hline
                         & F1     & Recall & Precision & Accuracy \\ \hline
GPT-4                      & 0.6809                     & \textbf{1}                     & 0.5162                       & 0.5162    \\
GPT-3.5                   & 0.6809                     & \textbf{1}                     & 0.5162                       & 0.5162                       \\ %
CodeLlama-13b           & 0.6767                      & 0.9781                      & 0.5173                         & 0.5176                        \\ %
CodeLlama-34b            & 0.6725                      & 0.9454                      & 0.5219                         & 0.5247                        \\ \hline
\approach & \textbf{0.9121} & 0.8934 & \textbf{0.9316}    & \textbf{0.9111}   \\ \hline
\end{tabular}}
\end{table}

\begin{table}[t!]
\centering
\caption{Performance comparison between \approach's Detector and other fine-tuned models. }
\label{tab:fine_tuning_compare}
\large
\scalebox{0.75}{
\begin{tabular}{l|c|c|c|c}
\hline
                         & F1     & Recall & Precision & Accuracy \\ \hline
CodeBERT                 & 0.8221 & 0.7322 & 0.9371    & 0.8364   \\ %
GraphCodeBERT            & 0.8841 & 0.8333 & 0.9414    & 0.8872   \\ %
CodeT5                   & 0.8481 & 0.7705 & \textbf{0.9431}    & 0.8575   \\ %
UnixCoder                & 0.8791 & 0.8443 & 0.9169    & 0.8801   \\ %
CodeLlama-13b-class          & 0.8936 & 0.8716 & 0.9167    & 0.8928   \\ \hline
\approach & \textbf{0.9121} & \textbf{0.8934} & 0.9316    & \textbf{0.9111}   \\ \hline
\end{tabular}}
\end{table}

Secondly, we compared Detector with fine-tuned models in detecting vulnerabilities, using F1, recall, precision, and accuracy as evaluation metrics, as shown in \tabref{fine_tuning_compare}.
We compared our method with CodeBERT, GraphCodeBERT, CodeT5, UnixCoder, and CodeLlama-13b-class.
CodeBERT, GraphCodeBERT, CodeT5, and UnixCoder underwent full model fine-tuning.
These traditional pre-trained models use the first token of the input sequence as the feature input for the classifier.
CodeBERT is based on the transformer encoder.
GraphCodeBERT has the same architecture as CodeBERT but includes additional pre-training on data dependency relations.
CodeT5 utilizes the transformer encoder and decoder, adopting an architecture similar to T5~\cite{10.5555/3455716.3455856}.
UnixCoder unifies the encoder and decoder architecture, controlling the model behaviour through a masked attention matrix with prefix adapters.
CodeLlama-13b-class performs classification based on LoRA. We fine-tuned CodeLlama-13b-class for LoRA classification using the PEFT framework. CodeLlama-13b-class uses the representation of the last token of the input sequence as the feature input for the classifier.

As shown in \tabref{fine_tuning_compare}, \approach achieves the highest scores of F1, Recall, and Accuracy among all methods, $0.9121$, $0.8934$ and $0.9111$.
CodeLlama-13b-class is second only to our method regarding vulnerability detection rate, and the performance is relatively close. GraphCodeBERT and UnixCoder perform worse than CodeLlama-13b-class. Although CodeT5 achieves the highest precision at $0.9431$, its other metrics are lower than GraphCodeBERT and UnixCoder.  CodeBERT has the worst performance. Additionally, the accuracy scores of these models are relatively high (all are more than $0.91$), indicating that many of the predicted risky vulnerabilities are indeed risky.

Regarding precision, \name produce more false positives than CodeT5, GraphCodeBERT, and CodeBERT.
This is primarily because \name uses a longer context length (16k) compared to these models, whose 512-length context truncation causes the vulnerable and safe samples to align more closely in length, thus affecting the apparent code complexity. However, our dataset shows that vulnerable code is relatively more complex than safe code. This leads \name to output 10 unique false positives with more complex code.

\answer{1}{
The performance of \name's Detector exceeds that of traditional full-model fine-tuning, LoRA fine-tuning in a classification manner, and LLMs based on in-context learning. The performance of fine-tuned models is also better than that of zero-shot learning.
}

\subsection{RQ2 - Explanation Alignment}

\begin{figure}[t!]
    \centering
    \scalebox{0.9}{
    \includegraphics[width=0.5\textwidth]{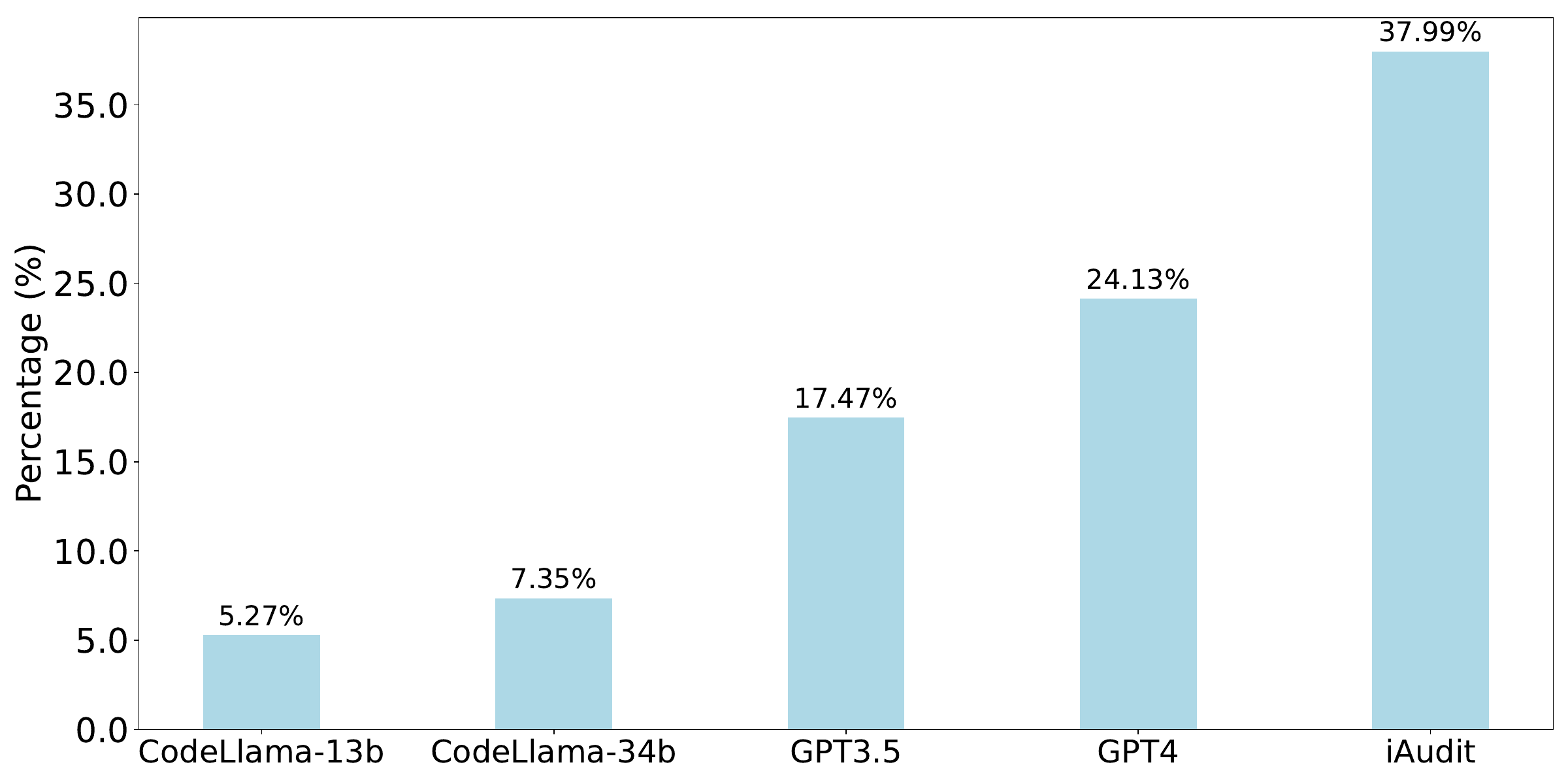}}
    \caption{Comparing the alignment with ground-truth reasons.} %
    \label{fig:root_cause_consisterncy}
\end{figure}

To measure the effectiveness of Reasoner in explaining vulnerabilities, we compared the consistency between the explanations we generated and the root causes.
Given the LLM's outstanding performance in interpreting textual meaning, we used GPT-4 to verify whether our generated explanations align with the root causes.
For this consistency assessment, we employed automated annotation prompts from Y. Sun et al.~\cite{sunLLM4VulnUnifiedEvaluation2024a}.
The results of our consistency test are depicted in \figref{root_cause_consisterncy}, where the y-axis represents the percentage of our explanations in the test set that match the root causes.
Our method significantly outperformed the baseline methods, achieving a consistency rate of 37.99\%, while no baseline method exceeded 25\%.
Among these baselines, GPT-4 performed the second best with 24.13\% consistency.
Additionally, the results also indicated that CodeLlama-13b had the weakest performance.

\answer{2}{
The rationality of Reasoner's output is clearly superior to that of other models. On the test set, its consistency with real reasons reaches 37.99\%, which is over 10\% higher than the second-ranked GPT-4.
}

\subsection{RQ3 - Two-stage Approach vs. An Integration Model}

\begin{figure}[t!]
    \centering
    \scalebox{0.9}{
    \includegraphics[width=0.5\textwidth]{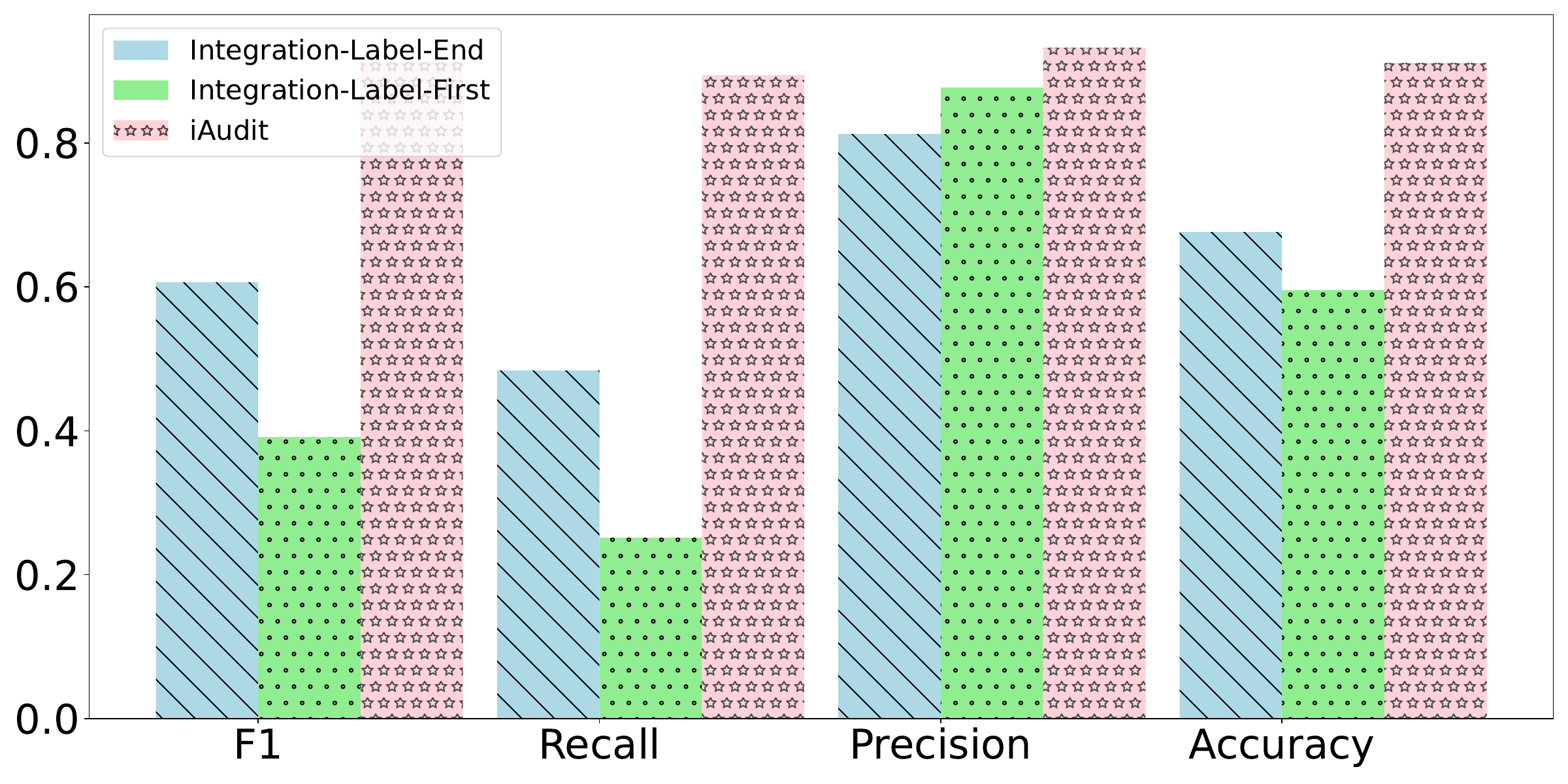}}
    \caption{Comparing \name with the integration models that make decisions and explain the vulnerabilities simultaneously.}
    \label{fig:integration}
\end{figure}

Our research methodology involves vulnerability detection and explanation, executed in two stages.
We trained two models, i.e., Detector and Reasoner, based on a generative approach to perform these tasks on their respective high-quality datasets.
A question arises whether these two tasks can be merged and trained simultaneously in a single model.
In response, we developed an integration model that generates labels and explanations for the vulnerabilities, comparing it to our two-stage approach.
The integration model uses prompts similar to those of Reasoner, with additional requirements to output the label.
We explored two integration training approaches: 1) generating labels first, then explaining the reasons; 2) explaining the reasons first, then generating labels.

The results, as shown in \figref{integration}, indicate that our step-wise approach outperforms the integration models across four key performance metrics.
While the three methods are similar in precision, the differences in other metrics are notable.
Analyzing these results, we found that the integration methods have higher accuracy for negative samples but a lower detection rate for positive samples (i.e., lower recall).
This may be attributed to the generative loss optimization, where the output sequence is longer, making the label-related loss occupy a smaller proportion of the total loss, thus preventing the model from adequately focusing on the label.
To test this hypothesis, we added data that includes only label generation to the dataset during the integration training process, guiding the model to focus more on the label.
In the evaluation phase, we still required the model to output both labels and explanations simultaneously.
Through this mixed training approach, we observed a significant improvement in the model's vulnerability detection performance, with an F1 score of $0.8433$, a recall rate of $0.8164$, a precision of $0.8723$, and an accuracy of $0.8434$.

\answer{3}{
\tool achieved better detection performance than the integration model that outputs labels and reasons simultaneously.
We confirmed that the model struggles to focus on the labels when required to output both types of information, as evidenced by our inclusion of label-only data in the verification process.
}

\begin{table}[h!]
\caption{Majority Voting vs. Single Prompt.}
\label{tab:majority_res}
\centering
\scalebox{0.8}{
\large
\begin{tabular}{l|c|c|c|c}
\hline
                         & F1     & Recall & Precision & Accuracy \\ \hline
Single-prompt & 0.8278 &  0.8005 & 0.8567 &  0.8279 \\ \hline
Prompt-1                 & 0.8988 & 0.8852 & 0.9127    & 0.8970   \\ %
Prompt-2                 & 0.9027 & 0.8743 & \textbf{0.9329}    & 0.9027   \\ %
Prompt-3                 & 0.9063 & 0.8852 & 0.9284    & 0.9055   \\ %
Prompt-4                 & 0.9098 & \textbf{0.8962} & 0.9239    & 0.9083   \\ %
Prompt-5                 & 0.9096 & 0.8934 & 0.9263    & 0.9083   \\ \hline
\approach & \textbf{0.9121} & 0.8934 & 0.9316    & \textbf{0.9111}   \\ \hline
\end{tabular}}
\end{table}

\subsection{RQ4 - Effectiveness of Majority Voting}

Our research explored a method using multiple prompts and a voting mechanism for Detector to determine the final label.
This method aims to enhance the model's precision and credibility.
During the evaluation process, we continued to use metrics such as the F1 score, recall, precision, and accuracy.
We calculated these metrics for each prompt individually for comparative analysis, as shown in \tabref{majority_res}.
It should be noted that the first row \textit{Single-prompt} indicates that we used only one prompt format to train Detector.
Prompt-1, Prompt-2, Prompt-3, Prompt-4, and Prompt-5 represent the results for each prompt after multiple-prompt training.
The last row shows the results after majority voting, indicating that majority voting can improve the overall performance of \approach, with both the F1 score and accuracy being the highest.
At the same time, except for Single-prompt, we noticed minimal performance differences among multiple prompts.
Single-prompt performed much worse than the others. Training with multiple prompts can improve model performance compared to using only one prompt during training.

\begin{figure}[t!]
    \centering
    \scalebox{0.9}{
         \includegraphics[width=0.5\textwidth]{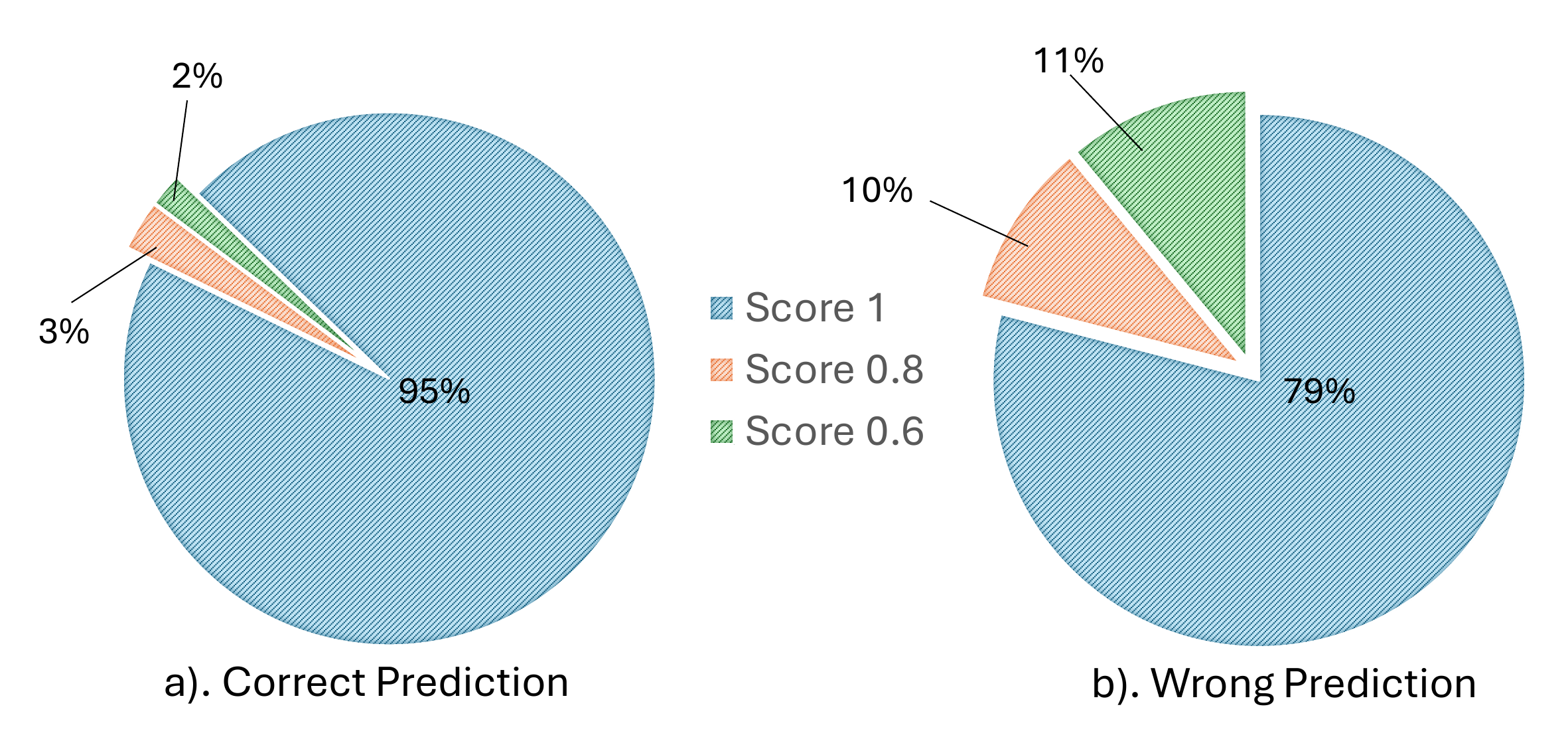}}
    \caption{The Distribution of Voting Scores for Correct Predictions and Wrong Predictions.}
    \label{fig:voting_dis}
\end{figure}

Additionally, we divided the test set into two groups based on whether the predictions were correct or incorrect, named ``correct prediction" and ``incorrect prediction" groups, respectively, and analyzed the distribution of confidence scores within these two groups.
We found that in the incorrect prediction group, the proportion of confidence scores within the range of $0.6$ to $0.8$ is significantly higher than in the correct prediction group (11\% vs 2\%, 10\% vs 3\%, respectively), as shown in \figref{voting_dis}.
The confidence score can reflect the reliability of the prediction results to a certain extent.
When the confidence score is low, the prediction results are less credible.

\answer{4}{
Majority voting enhances the detection performance and stability.
Additionally, using multiple prompts allows the model to perform better and be more reliable than when using a single prompt.
}

\subsection{RQ5 - Impact of Additional Information}

We explored whether introducing additional call graph information into the model could enhance its performance. We added function call relationships to the prompts as contextual information.

\paragraph{RQ5.1} Through comparative experiments as shown in \tabref{additional_info}, we found that this calling contextual information did not improve the model's overall performance.
In the second row, \textit{Call}, we used the prompts with the calling information and then employed majority voting to decide the prediction result.
For the third row, \textit{Call-OutCall}, we used prompts both with and without calling information and also used majority voting.
Compared with \tool's Detector, they exhibited lower precision, accuracy, and f1, with almost the same recall.

\begin{figure}[t!]
    \centering
    \scalebox{0.8}{
    \includegraphics[width=0.5\textwidth]{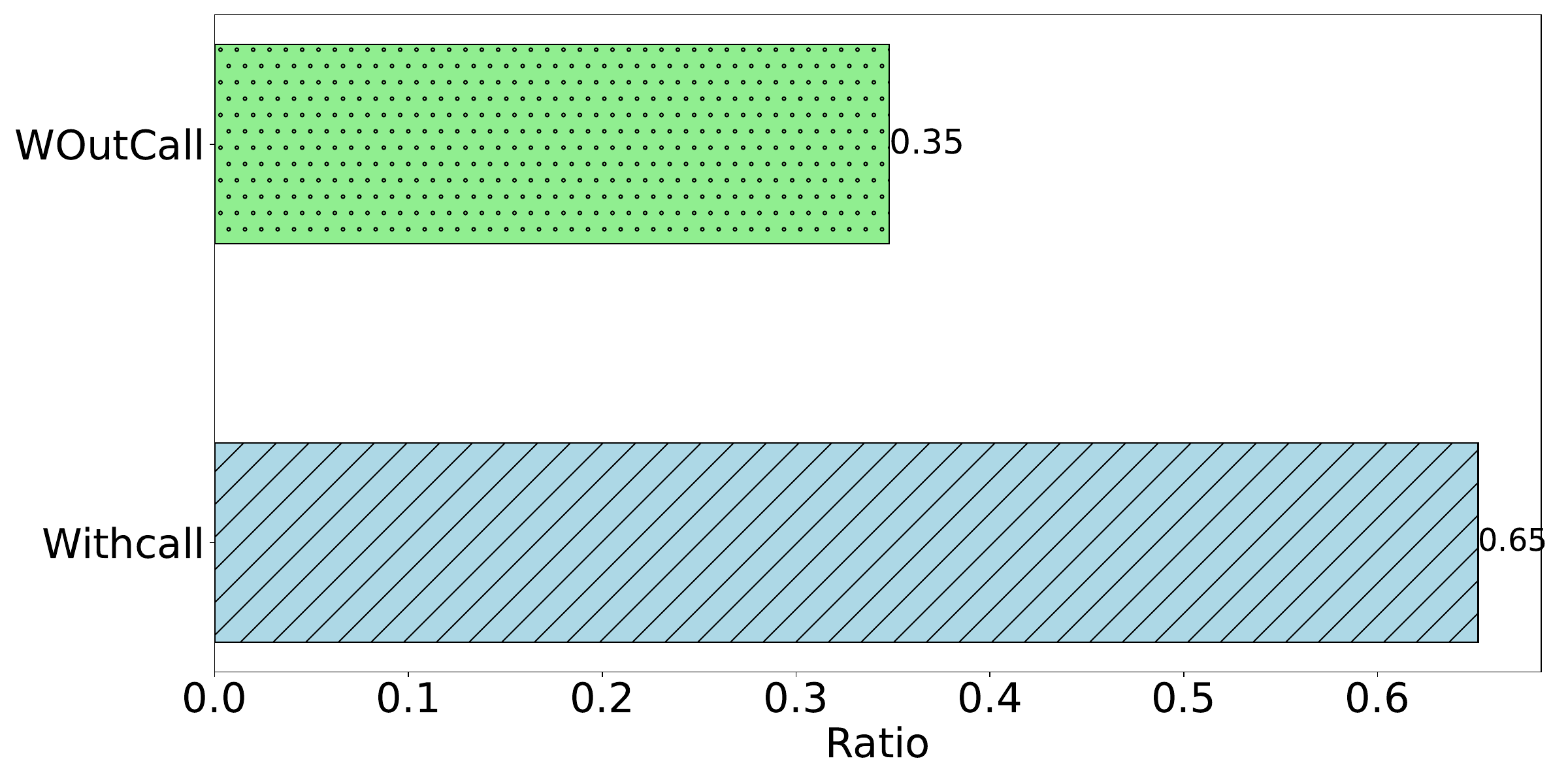}}
    \caption{Final Reason Distribution of Ranker-Critic.}
    \label{fig:reason_distribution}
\end{figure}

\paragraph{RQ5.2} \figref{reason_distribution} demonstrates the selected reason distribution from Ranker-Critic. We can see that the majority~(65\%) of the selected reasons are from the prompts with calling information, while there is still a high ratio~(35\%) of selected reasons from the prompts without calling information.

Although function call relationships provide more information, this information does not always help the model better complete the current task.
In some cases, this information may cause interference, making it difficult for the model to identify critical information, thereby resulting in more false positives and affecting performance.
Furthermore, not all function call relationships are practically valuable.
If these additional pieces of information are not closely related to the problem the model is trying to solve, they may not help enhance the model's performance.
Our research indicates that merely adding function call information does not directly facilitate the model's effectiveness in detecting vulnerabilities.
In the field of vulnerability detection, exploring how to construct effective contextual information remains a challenging and worthy research question.

\answer{5}{
Additional call graph information may enable the model to make better judgments in some cases.
However, we also observed situations where this additional information could potentially confuse the model, thereby reducing its performance.
}

\begin{table}[t!]
\caption{Impact with or without Additional Information.}
\label{tab:additional_info}
\centering
\scalebox{0.85}{
\large
\begin{tabular}{l|c|c|c|c}
\hline
         & F1     & Recall & Precision & Accuracy \\ \hline
Call     & 0.9011 & \textbf{0.8962} & 0.9061    & 0.8984   \\ %
Call-OutCall 
& 0.9083 & 0.8934 & 0.9237 & 0.9069 \\ \hline
\approach & \textbf{0.9121} & 0.8934 & \textbf{0.9316}    & \textbf{0.9111}   \\ \hline
\end{tabular}}
\end{table}

%% file: sections/related.tex
\section{Related Work}
\label{sec:related}

Vulnerability detection has been a critical issue for the healthy and sustainable development of the software ecosystem, especially in blockchain and smart contracts.
Traditional vulnerability detection methods, such as those based on predefined static analysis rules~\cite{feistSlitherStaticAnalysis2019}, often lack robust generalization capabilities and are difficult to adapt to new types of vulnerabilities.
Moreover, some logic-related vulnerabilities~\cite{zhangDemystifyingExploitableBugs2023b} are also challenging to encapsulate into static analysis rules.
To address this issue, researchers have employed deep learning-based approaches.
For example, Zhuang et al.\cite{zhuangSmartContractVulnerability2020} used graph neural networks to detect vulnerabilities in smart contracts. Liu et al.\cite{liuSmartContractVulnerability2021} combined interpretable graph features with expert patterns to achieve better results and interpretable weights.
Wu et al.~\cite{wuPeculiarSmartContract2021} utilized a pre-training technique and critical data flow graphs for the detection of smart contract vulnerabilities.

With the advent of large language models, researchers are not only utilizing traditional deep learning models but also LLMs for vulnerability detection.
For example, Ullah et al.~\cite{ullahCanLargeLanguage2023b} evaluated LLMs on vulnerability detection tasks and found that they may not perform well.
Fu et al.~\cite{fuChatGPTVulnerabilityDetection2023a} further analyzed the gap for LLMs in detecting vulnerabilities.
Thapa et al.~\cite{thapaTransformerBasedLanguageModels2022b} leveraged LLMs for software vulnerability detection, and David et al.~\cite{davidYouStillNeed2023c} used LLMs for smart contract vulnerability tasks.
Alqarni et al.~\cite{alqarniLowLevelSource2022} fine-tuned the BERT model~\cite{devlinBERTPretrainingDeep2019b} for source code vulnerability detection.
Sun et al.~\cite{sunLLM4VulnUnifiedEvaluation2024a} proposed an unified evaluation framework, LLM4Vuln, to enhance the ability of LLMs to detect vulnerabilities.
Some research also fused large language models with traditional program analysis methods.
Sun et al.~\cite{sunGPTScanDetectingLogic2023c} proposed GPTScan for smart contracts, leveraging static program analysis to reduce the false positives of LLMs.
Li et al.~\cite{liHitchhikerGuideProgram2023a} proposed LLift for integrating LLMs with static analysis tools.
SmartInv~\cite{wang2024smartinv} and PropertyGPT~\cite{liu2024propertygpt} further integrated large language models with formal verification methods, aiming not only to detect vulnerabilities but also to prove that a piece of code is secure.

However, all these studies have not tuned domain-specific knowledge into the models themselves, focusing only on the knowledge from the pre-training dataset or the vulnerable code segment itself, which could not effectively detect logic bugs.

%% file: sections/threats.tex
\section{Threats to Validity}
\label{sec:threats}

In the data collection process, there is a risk of data bias, which might prevent models trained and tested on these data from generalizing accurately.
Moreover, the precision of data labelling significantly impacts model performance.
To mitigate these issues, we collected verified data from real and public audit reports and utilized the latest tools, such as GPTScan~\cite{sunGPTScanDetectingLogic2023c} and LLM4Vuln~\cite{sunLLM4VulnUnifiedEvaluation2024a}, to assist in cleaning and annotating the data.
It is important to note that external links in the data could induce LLM to produce incorrect information; therefore, we performed data cleaning to remove these links.
Considering LLMs' sensitivity to input data, we standardized the code data, including removing unnecessary spaces without changing code semantics, to enhance the model robustness and reliability.
To maximize the performance of the zero-shot learning of GPT-3.5 and GPT-4, we adopted and optimized the prompts from our partner, MetaTrust Labs.
These prompts have been integrated into their working pipeline.
For open-source models, we collaborated with an auditing expert to adapt their prompts for these models.

Overfitting is a common issue during model training, which we addressed by implementing an early stopping strategy.
The choice of different models might affect the ranker-critic architecture's effectiveness.
We tested multiple cutting-edge open-source models, including MoE~\cite{jiang2024mixtral}, CodeLlama-70b~\cite{roziere2023code}, Llama2-70b~\cite{touvron2023llama}, and the recently introduced Gemma~\cite{team2024gemma}, and compared their performance on inference benchmark tests.
Based on factors like the strictness of the model output format and operational speed, we chose MoE~\cite{jiang2024mixtral}.
Our research also showed that the consistency between the selected reasons from the MoE and the real reasons reached about 38\%.
To control costs, we limited the maximum iterations in the ranker-critic loop to five and adopted four-decimal precision handling.

%% file: sections/limitation.tex
\section{Limitation}
Although our method performed excellently in trials, its primary advantage lies in detecting logic vulnerabilities in smart contracts, which account for more than 80\% of exploitable vulnerabilities according to a recent study~\cite{zhangDemystifyingExploitableBugs2023b}. As such, while \name's training data does include instances of Reentrancy, Overflow, and Underflow vulnerabilities, handling them is not the major usage scenario of \name. Indeed, these traditional contract vulnerabilities are theoretically more suitable for detection by program analysis methods. That said, both AI-based and PL-based methods have their unique comfort zones, and since \name is fully AI-based, this paper compared it with other AI-based methods only. Additionally, despite our efforts to mitigate the phenomenon of hallucinations in large language models (LLMs) through voting and agents, hallucinations may still occur. Finally, since our tools rely on LLMs, there are certain hardware requirements. While the models can be compressed to run on less powerful GPUs, this compression may result in reduced model performance.

%% file: sections/conclusion.tex
\section{Conclusion}
\label{sec:conclusion}
In this paper, we proposed \tool, the first smart contract auditing framework that combines fine-tuning and LLM-based agents to detect vulnerabilities and explain the results.
We adopted a multiple-prompt-based strategy and applied LoRA-based fine-tuning to train the Detector and Reasoner. The former generates results based on a majority voting mechanism, while the latter provides multiple alternative explanations based on different inference paths.
Furthermore, we introduced two LLM agents, Ranker and Critic, to collaborate in selecting the most appropriate explanation. 
Our approach demonstrated superior performance in zero-shot scenarios compared to zero-shot LLM learning and traditional full-model fine-tuning methods.
We studied the performance improvement brought by the majority voting strategy and compared different LoRA training methods, providing the rationality of our choice.
We also explored how additional calling context affects our model's performance. 
For future work, we will focus on enhancing the model's stability and its alignment with human preferences.